\documentclass[a4paper]{jpconf}
\usepackage{graphicx}
\usepackage{array}
\usepackage{float}
\usepackage{amssymb}
\usepackage{color}

\begin{document}
\title{Type I Seesaw Mechanism, Lepton Flavour Violation and Higgs Decays}

\author{Emiliano Molinaro}

\address{Physik-Department T30d, Technische 
Universit\"at M\"unchen,\\ James-Franck-Stra{\ss}e 1, 85748 Garching, Germany.\\}

\ead{emiliano.molinaro@tum.de}

\begin{abstract}
We review and update the current phenomenological constraints on minimal type I seesaw extensions of the Standard Model in which  New Physics content can be probed at the electroweak scale. In this class of models, the flavour structure of the neutrino Yukawa couplings is determined by the requirement of reproducing neutrino oscillation data. The strongest constraints on the seesaw parameter space are imposed by the very recent upper limit on $\mu\to e\,\gamma$ decay rate. Searches of non-standard Higgs boson decays into a light and a heavy neutrino may also provide and independent test of these seesaw scenarios.
\end{abstract}

\section{Introduction}

The recent discovery of the Higgs boson at the Large Hadron Collider (LHC) \cite{Aad:2012tfa, Chatrchyan:2012ufa} with a mass $m_h\sim 125$ GeV marks an important breakthrough in our understanding of the mechanism of electroweak (EW) symmetry  breaking and is by itself  an extraordinary success of the Standard Model (SM) of Elementary Particles.

On top of the discovery of the Higgs boson, there are still several open questions which the SM, in the way it is conceived  now, cannot explain. In particular, New Physics must be advocated to explain the outcome of the flavour neutrino oscillation experiments, which provide compelling and indubitable evidence that at least two neutrinos have non-zero masses, several orders of magnitude smaller than the other SM charged fermions masses, and that the three neutrino flavours mix. 
These very small neutrino masses can be linked to a new physical scale of Nature that, in general, is not related to the EW symmetry breaking scale. 

One of the most successful mechanism of neutrino mass generation is the well known type I seesaw scenario \cite{seesaw}, in which the SM is extended with at least two
heavy SM singlet fermions, $N_{1,2}$, usually dubbed right-handed neutrinos, that have Yukawa-type interactions with the SM Higgs and left-handed doublets. If further symmetries are not imposed, such neutrino Yukawa couplings explicitly break lepton number and a Majorana mass term for the SM neutrinos is generated at tree-level after the decoupling of the heavy fermions $N_{1,2}$. 

In the following we consider the type I seesaw scenarios studied in  \cite{Ibarra:2010xw,Ibarra:2011xn,Dinh:2012bp,Cely:2012bz} in which the right-handed neutrino mass scale is set in the TeV range, that is $M_{1,2}\sim(100-1000)$ GeV.
This class of models predicts a rich low energy and collider phenomenology and, for this reason, the full seesaw parameter space is tightly constrained by different sets of data: 
neutrino oscillation data, neutrinoless double beta decay, EW precision tests and charged lepton flavour violating processes and right-handed neutrino production/detection at colliders.

As discussed below, the unknown seesaw parameter space can be conveniently expressed in terms of the size of the neutrino Yukawa couplings ($y$) and the heavy Majorana neutrino masses ($M_{1,2}$). The most conservative bound on a combination of these parameters is set by current experiments searching for $\mu\to e\,\gamma$ decay. 

In the case of $M_{1,2} \lesssim m_h$
it is also possible to probe the neutrino Yukawa interactions looking for non-standard Higgs boson decays into a light and a heavy Majorana neutrino \cite{Dev:2012zg,Cely:2012bz,de Gouvea:2007uz,Pilaftsis:1991ug,Chen:2010wn}.\\

All the present limits on the seesaw parameter space are summarized in Figs~\ref{fig1} and \ref{fig2}.

\section{Constraints on the seesaw parameter space}

The mixing between the  new singlet SM heavy Majorana fermions $N_{1,2}$ and the SM light neutrinos generate charged and neutral current interactions between $N_{1,2}$ and the SM gauge/Higgs bosons, which can in principle be tested in low energy and collider experiments for masses $M_{1,2}\sim(100-1000)$ GeV. 
The relevant parts of the interaction Lagrangian are in this case
\begin{eqnarray}
 \mathcal{L}_{CC}^N &=& -\,\frac{g}{2\,\sqrt{2}}\,
\bar{\ell}\,\gamma_{\alpha}\,(RV)_{\ell k}(1 - \gamma_5)\,N_{k}\,W^{\alpha}\;
+\; {\rm h.c.}\,
\label{NCC},\\
 \mathcal{L}_{NC}^N &=& -\frac{g}{4 \,c_{w}}\,
\overline{\nu_{\ell L}}\,\gamma_{\alpha}\,(RV)_{\ell k}\,(1 - \gamma_5)\,N_{k}\,Z^{\alpha}\;
+\; {\rm h.c.}\,,\label{NNC}\\
\mathcal{L}_{H}^N &=& -\frac{g M_{k}}{4\, M_{W}}\,
\overline{\nu_{\ell L}}\,(RV)_{\ell k}\,(1 + \gamma_5)\,N_{k}\,h\;
+\; {\rm h.c.}\,
\label{NH}
\end{eqnarray}
The copulings $(RV)_{\ell k}$ ($\ell=e,\mu,\tau$ and $k=1,2$) arise from the mixing between heavy and light Majorana neutrinos and, therefore, are suppressed by the seesaw scale. They can be conveniently parametrized as follows~\cite{Ibarra:2011xn}:
 \begin{eqnarray}
\label{mixing-vs-y}
\left|\left(RV\right)_{\ell 1} \right|^{2}&=&
\frac{1}{2}\frac{y^{2} v^{2}}{M_{1}^{2}}\frac{m_{3}}{m_{2}+m_{3}}
    \left|U_{\ell 3}+i\sqrt{m_{2}/m_{3}}U_{\ell 2} \right|^{2}\,,
~~{\rm NH}\,,\\
\left|\left(RV\right)_{\ell 1} \right|^{2}&=&
\frac{1}{2}\frac{y^{2} v^{2}}{M_{1}^{2}}\frac{m_{2}}{m_{1}+m_{2}}
    \left|U_{\ell 2}+i\sqrt{m_{1}/m_{2}}U_{\ell 1} \right|^{2}
\simeq \;\frac{1}{4}\frac{y^{2} v^{2}}{M_{1}^{2}}
\left|U_{\ell 2}+iU_{\ell 1} \right|^{2}\,,
\,{\rm IH}\,,
\label{mixing-vs-yIH}\\
(RV)_{\ell 2}&=&
\pm i\, (RV)_{\ell 1}\sqrt{\frac{M_1}{M_2}}\,,~\ell=e,\mu,\tau\,,
\label{rel0}
\end{eqnarray}
%%%%%%%%%%%%%%%%%%%%%%%%%%%%
%
where $U$ denotes the neutrino mixing matrix and $v\simeq174$ GeV. 
Notice that the relative mass splitting of the two heavy Majorana neutrinos 
must be very small, $|M_{1}-M_{2}|/M_{1}\ll 1$, due to the current upper limit on the effective Majorana mass probed in neutrinoless double beta decay experiments \cite{Ibarra:2010xw,Ibarra:2011xn}. In this case, the flavour structure of the neutrino Yukawa couplings is fixed by the neutrino oscillation parameters \cite{Ibarra:2010xw,Ibarra:2011xn,Raidal:2004vt} and the two heavy Majorana neutrinos form a pseudo-Dirac fermion: $N=(N_1\pm i N_2)/\sqrt{2}$. The parameter $y$ in the expressions above represents the largest
eigenvalue of the matrix of the neutrino Yukawa couplings
\cite{Ibarra:2011xn}:
%%%%%%%%%%%%%%%%%%%%%%%%%%%%%%%%%
\begin{equation}
\label{ymax2}
y^{2}v^{2}\;=\;2\,M_{1}^{2}\,\left(\left| (RV)_{e1} \right|^{2}+
\left| (RV)_{\mu1} \right|^{2}+\left| (RV)_{\tau1} \right|^{2}\right)\,.
\end{equation}  
%%%%%%%%%%%%%%%%%%%%%%%

\subsection{Charged lepton flavour violation: $\mu\to e\, \gamma$}
%%%%%%%%%%%%%%%%%%%%%%%%%%%%%%%
\begin{figure}
\begin{center}
\begin{tabular}{cc}
\includegraphics[width=7.5cm,height=6.5cm]{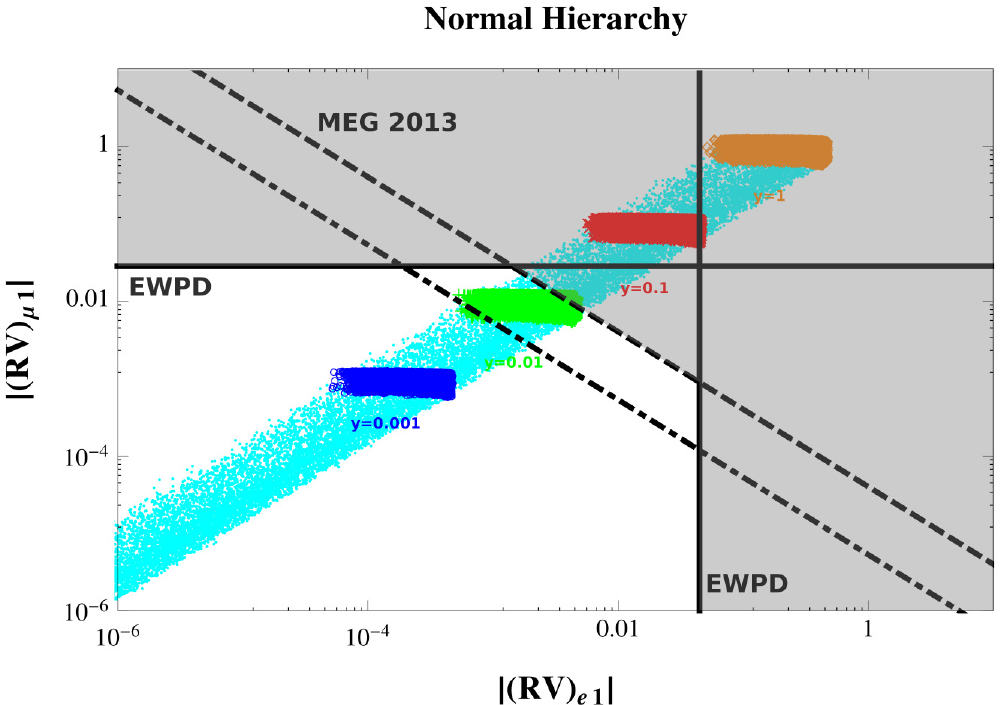} &
\includegraphics[width=7.5cm,height=6.5cm]{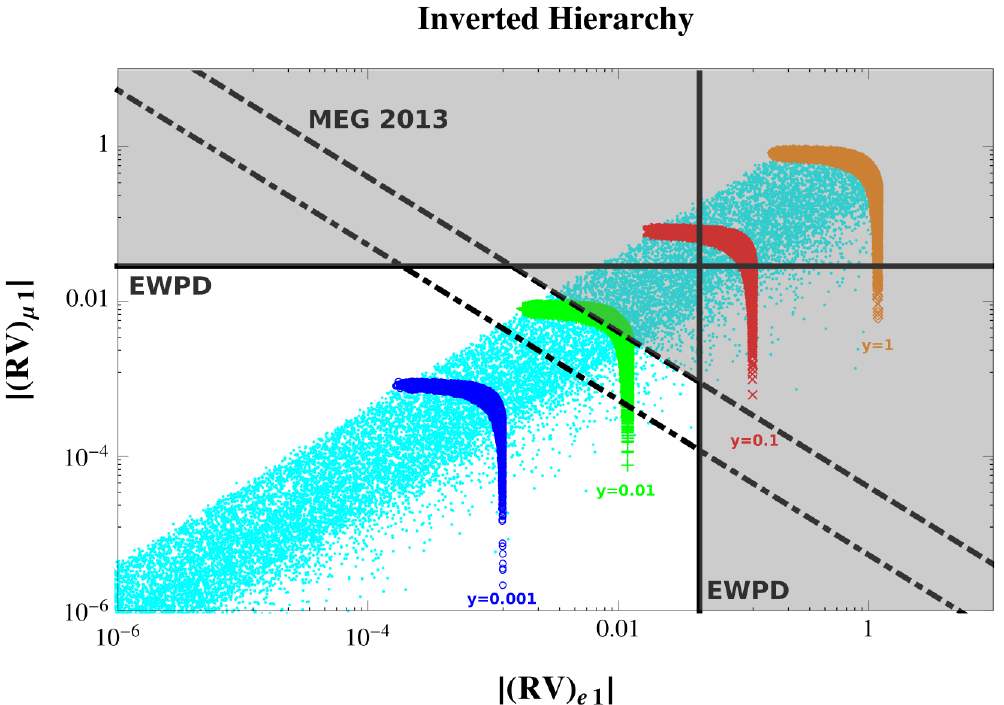}
\end{tabular}
\caption{Correlation between $|(RV)_{e1}|$ and $|(RV)_{\mu1}|$ for 
$M_{1}=100$ GeV in the case of normal (left panel)
and inverted (right panel) light neutrino mass spectrum. The regions corresponding to four different values of the neutrino Yukawa eigenvalue $y$ are highlighted.
The cyan points correspond
to random values of $y\leq 1$. The dashed line represents to the present MEG
bound \cite{MEG}, $B(\mu\rightarrow e \,\gamma)\leq
5.7 \times 10^{-13}$. The dot-dashed line
coincides with  $B(\mu\rightarrow e \,\gamma) =
10^{-14}$.\label{fig1}}
\end{center}
\end{figure}
%%%%%%%%%%%%%%%%%%%%%%%%%%%%%%%%%%%%

In the case of a seesaw mass scale $M_{1,2}$ in the TeV range considered here, the most stringent constraint on the size of the neutrino Yukawa couplings is set by the very recent upper limit on $\mu^+\to e^+\, \gamma$ branching ratio reported by the MEG Collaboration \cite{MEG}: ${\rm BR}(\mu\to e\, \gamma)<5.7\times 10^{-13}$ at 90\% confidence level.
Taking the best fit values of the neutrino oscillation 
parameters \cite{Tortola:2012te}, we can estimate the typical size of neutrino Yukawa couplings:
\begin{eqnarray}
&& y\lesssim 0.026\,~~~{\rm for~~~} M_1\;=\;100\,{\rm GeV}\,.
\label{yupNH}
\end{eqnarray}

We implement in Fig.~\ref{fig1} all the constraints on the effective couplings $(RV)_{\mu 1}$ and $(RV)_{e1}$ which come from the requirement of reproducing neutrino oscillation data, from EW precision data and from the current upper limit on $\mu\to e\,\gamma$. The seesaw mass scale is fixed at benchmark value: $M_1=100$ GeV.
From Fig.~\ref{fig1}, it is manifest that the
allowed ranges of the right-handed
neutrino couplings $|(RV)_{\mu1}|$ and $|(RV)_{e 1}|$, 
in the case of normal (left panel) and inverted (right panel) light neutrino mass spectrum, are confined in a small strip of the overall representative plane. 
This corresponds to the scatter plot of the points 
which are consistent with the $3\sigma$ allowed
ranges of the neutrino oscillation parameters \cite{Tortola:2012te}.
The region of the parameter space 
which is allowed by the EW precision data,
is marked with solid lines. The region allowed
by the current bound on the $\mu\rightarrow e\,\gamma$ 
decay rate is indicated with a dashed line, while the dot-dashed line shows the projected MEG sensitivity reach, $i.e.$ ${\rm BR}(\mu\to e\,\gamma)\sim 10^{-14}$. 
The scatter points corresponds to different values of 
the maximum neutrino Yukawa coupling $y$: 
$y=0.001$ (blue $\circ$), $ii)$
$y=0.01$ (green $+$), $iii)$ $y=0.1$ (red $\times$), $iv)$ $y=1$ (orange $\Diamond$) and 
$v)$  an arbitrary value of the Yukawa coupling $y \leq 1$ (cyan points).

As depicted in Fig.~\ref{fig1}, in the case of a light neutrino mass spectrum with inverted hierarchy a strong suppression of the $\mu\to e\,\gamma$ decay rate is possible for specific values of the measured neutrino parameters.  Indeed, using the standard
parametrisation of the neutrino mixing matrix $U$ one can show that 
for fixed values of the mixing angles
$\theta_{12}$, $\theta_{23}$ and of the Dirac ($\delta$) and Majorana ($\alpha_{21}$) phases, $|U_{\mu 2}+iU_{\mu 1}|^{2}$
has a minimum for \cite{Ibarra:2011xn, Dinh:2012bp}
%%%%%%%%%%%%%%%%%%%%%%%%%%%%%%%%%%%%
\begin{eqnarray}
\sin\theta_{13} &=& \frac{c_{23}}{s_{23}}\,
\frac{\cos2\theta_{12} \cos\delta\sin\frac{\alpha_{21}}{2} -
\cos\frac{\alpha_{21}}{2}\sin\delta}
{1 + 2c_{12}\, s_{12}\, \sin\frac{\alpha_{21}}{2}}\,.
\label{s13min}
\end{eqnarray}
%%%%%%%%%%%%%%%%%%%%%%%%%%%%%%%%%%%%%
%
At the minimum 
%%%%%%%%%%%%%%%%%%%%%%%%%%%%%%%%%%%%
\begin{eqnarray}
{\rm min}\left(|U_{\mu 2}+i\,U_{\mu 1}|^{2}\right)&=&
c^2_{23}\, \frac{\left(\cos\delta\cos\frac{\alpha_{21}}{2} +
\cos2\theta_{12}\sin\delta\sin\frac{\alpha_{21}}{2}\right)^{2}}
{1 + 2c_{12}\, s_{12}\,\sin\frac{\alpha_{21}}{2}}\,.
\label{Umu21IHmin}
\end{eqnarray}
%%%%%%%%%%%%%%%%%%%%%%%%%%%%%%%%%%%
%
Therefore, the $\mu \to e\gamma$ branching ratio, that is proportional to eq.~(\ref{Umu21IHmin}), is
highly suppressed if the Dirac and Majorana phases take CP conserving values, mainly: $\delta\simeq 0$ and $\alpha_{21}\simeq \pi$. 
In this case, from eq.~(\ref{s13min}) we get a lower bound on the reactor angle, $i.e.$ $\sin(\theta_{13})\gtrsim 0.13$, which is in agreement with the global fit results \cite{Tortola:2012te}.
 On the other hand, assuming CP violating phases,
 ${\rm min}(|U_{\mu 2}+iU_{\mu 1}|^{2})\approx 0$ is possible, provided
$\theta_{12}$ and the Dirac and Majorana phases
$\delta$ and $\alpha_{21}$ satisfy the
following conditions:
$\cos\delta\cos(\alpha_{21}/2) +
\cos2\theta_{12}\sin\delta\sin(\alpha_{21}/2) \approx 0$
and ${\rm sgn}(\cos\delta\cos\frac{\alpha_{21}}{2})=
-{\rm sgn}(\sin\delta\sin\frac{\alpha_{21}}{2})$.

We remark that the upper limit given in (\ref{yupNH})  
is of the same order as the 
bottom Yukawa coupling, $y_{b}=m_{b}/v\simeq 0.024$.
Therefore it is interesting to explore the possibility to produce (pseudo-Dirac)
heavy neutrinos $N$
at present and forthcoming collider facilities  through non-standard Higgs boson decays, mainly $h\to \nu_{\ell L} + \overline{N},\;\overline{\nu_{\ell L}} + N$.

\subsection{New Higgs decay channel: $h\to \nu\, N$}

\begin{figure}[t]
\begin{center}
\includegraphics[width=16cm,height=10cm]{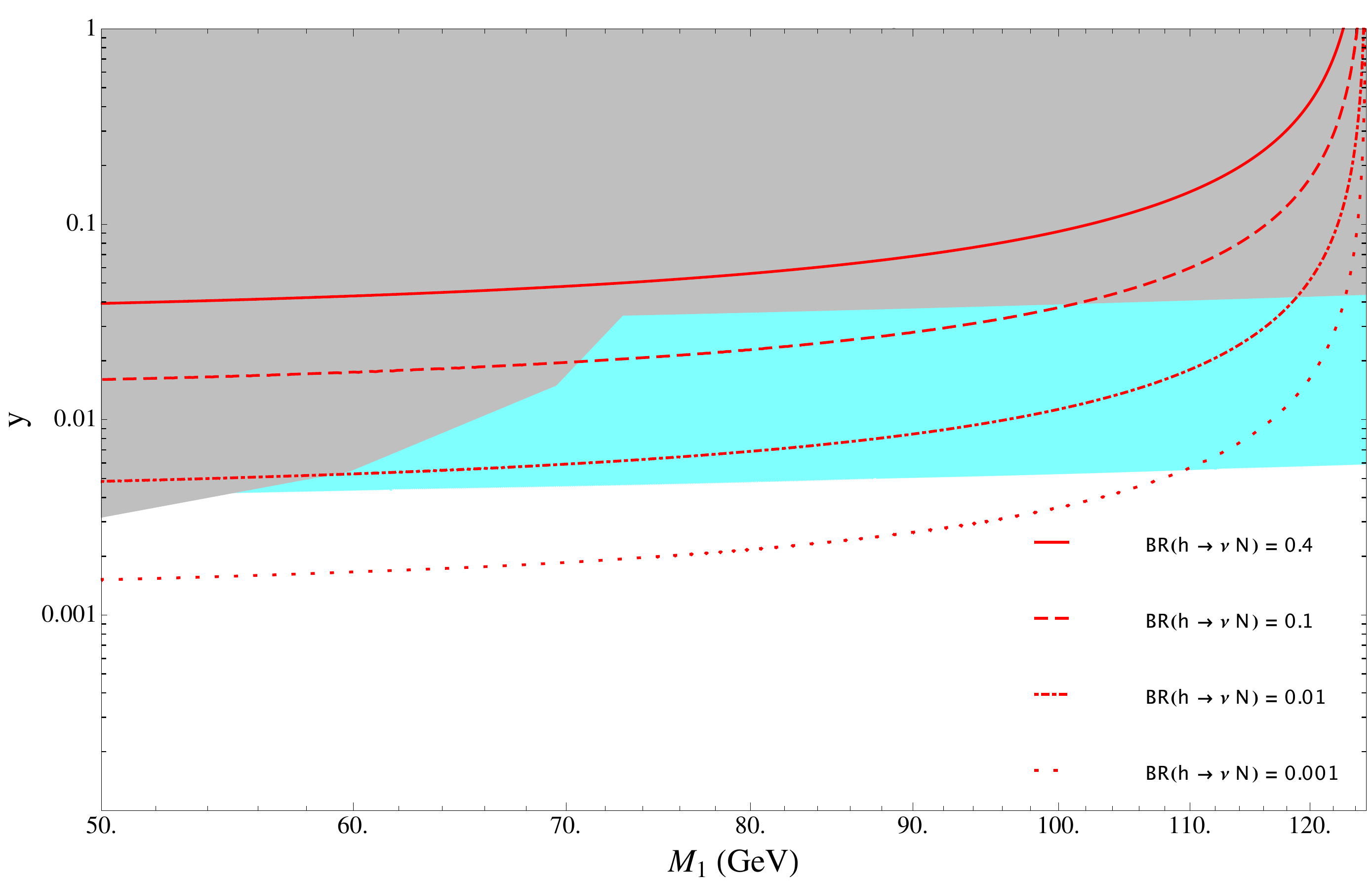}
\caption{Values of the maximum neutrino Yukawa coupling $y$ probed 
by Higgs decays into $N$ for $m_{h}=125$ GeV. 
The grey region is excluded by LEP2 data \cite{LEP2} and  
searches of lepton flavour violation \cite{MEG}. 
The cyan area represents the 
region of the parameter space which can be probed by 
the MEG experiment with the projected sensitivity 
to ${\rm BR}(\mu \rightarrow e \gamma) = 10^{-14}$.
}
\label{fig2}
\end{center}
\end{figure}

In the case of $M_1\lesssim m_h\sim 125$ GeV, it is possible to put limits on the seesaw parameter space looking for production of the right-handed neutrinos in non-standard Higgs boson decays: $h\to \nu\, N$ (see $e.g.$ \cite{Dev:2012zg,Cely:2012bz,de Gouvea:2007uz,Pilaftsis:1991ug,Chen:2010wn}).
In this case, the Higgs decay rate
is directly related to the neutrino Yukawa coupling $y$ 
defined in eq.~(\ref{ymax2}):
%%%%%%%%%%%%%%%%%%%%%%%%%%%%%%%%%%
\begin{eqnarray}
	\Gamma(h\to\nu N) &\equiv&\sum_{\ell=e,\mu,\tau}\,  \left( \Gamma(h\to \nu_{\ell L}\, \overline{N})+
	\Gamma(h\to \overline{\nu_{\ell L}}\,N) \right) \nonumber\\
	& = & 
\frac{1}{16\pi}\, y^{2}\, m_{h}\,
\left( 1-\frac{M_{1}^{2}}{m_{h}^{2}} \right)^{2}\nonumber\,.
\end{eqnarray}
%%%%%%%%%%%%%%%%%%%%%%%%%%%%%%%%
%
Taking as benchmark values $m_{h}=125$ GeV and $M_{1}=100$ GeV, 
one gets  $\Gamma(h\to\nu N)/\Gamma(h\to b\overline{b}) 
\simeq 0.19~ (y/0.05)^{2}$, that turns out to be sizable
if the upper limit on the Yukawa coupling $y$,  
obtained using the MEG upper bound (\ref{yupNH}) is saturated. 
Conversely, the search for the Higgs decay $h\to \nu N$ may provide
limits on the parameters of the low scale seesaw scenario which are
competitive to those from the searches for the $\mu\to e\,\gamma$ decay, 
when  $m_h>M_1$ \cite{Cely:2012bz}. On the other hand, in the case $M_1>m_h$ 
the exotic Higgs decay channels are,
$h\to\nu N\to \nu\, \nu\, Z,\; \nu\, \ell\, W$ 
which have a rate suppressed by the fourth power of $y$ 
as well as by the three-body 
decay phase space.

In Fig.~\ref{fig2} it is shown the correlation between $y$ and the seesaw scale $M_{1}$, corresponding to different 
values of ${\rm BR}(h\rightarrow \nu N)$ and the Higgs boson mass $m_{h}=125$ GeV. 
The  region in grey is excluded by the combined data on $\mu\rightarrow e\,\gamma$ decay
 \cite{MEG} and the search for heavy singlet neutrinos in 
$Z$ boson decays at LEP2~\cite{LEP2}.
It follows from the plot that the present bounds 
on the low scale seesaw parameter space do not 
preclude the possibility of a Higgs boson 
decaying into a heavy and a light neutrino 
with a branching ratio of roughly 10\%. 
 We also show in the plot 
as a cyan area the projected sensitivity reach 
of the MEG experiment searching 
for the $\mu\to e\,\gamma$ decay 
with a branching ratio ${\rm BR}(\mu\to e\,\gamma) \gtrsim 10^{-14}$,
which may allow to exclude 
${\rm BR}(h\to\nu N)\gtrsim 1\%$ for $M_{1}\gtrsim 100$ GeV.  

\section{Conclusions}

Minimal type I seesaw extensions of the Standard Model with right-handed neutrino masses 
at the EW scale predict by a very rich phenomenology and are tightly constrained by both low energy observables and collider searches. In the minimal scenario only two new fermion representations $N_{1,2}$ are introduced, which are Standard Model singlets and  have Yukawa-type couplings to the Higgs and lepton doublets, usually referred to as neutrino Yukawa couplings. The flavour structure of such interactions is univocally fixed by the requirement of reproducing the data on the neutrino masses and mixing. 
The existing low energy constraints imply that the two heavy right-handed neutrinos form a pseudo-Dirac pair, $N\equiv(N_1 \pm i N_2)/\sqrt{2}$, thus preventing the possibility to observe a possible signature of lepton number violation in this class of scenarios \cite{Ibarra:2010xw,Ibarra:2011xn}.
The phenomenological relevant seesaw parameter space can be conveniently expressed in term of the seesaw mass scale $M_{1}\approx M_2$ and the largest neutrino Yukawa eigenvalue $y$.

All the constraints on $y$ and $M$ are reported in Figs~\ref{fig1} and \ref{fig2}.
The strongest bounds are derived from the very recent upper limit on $\mu\to e\,\gamma$
decay rate  released by the MEG experiment \cite{MEG}: ${\rm BR}(\mu\to e\, \gamma)<5.7\times 10^{-13}$ at 90\% confidence level.
If the MEG experiment eventually observes the 
$\mu\to e\gamma$ decay with a branching ratio
${\rm BR}(\mu\to e\gamma) > 10^{-14}$,
the low scale type I see-saw scenario may be directly 
tested at LHC through non-standard Higgs boson decays $h\to\nu\,N$ decays \cite{Cely:2012bz}. 
On the other hand, if no positive signal is detected 
in the MEG experiment,
this will lead to a more stringent limit on the 
neutrino Yukawa coupling $y$ that will exclude for the moment
the possibility of
producing and detecting the new heavy 
pseudo-Dirac neutrino $N$.

\subsection*{Acknowledgments}
I would like to thank  Camilo Garcia Cely, Dinh N. Dinh, Alejandro Ibarra and
Serguey T. Petcov for their collaboration on the research discussed here.
This work is supported by the ERC Advanced Grant project ``FLAVOUR" (267104).

\end{document}